\documentclass[pra,twocolumn,,amsmath,amssymb,floatfix]{revtex4}
\usepackage{graphicx}
\usepackage{bm}
\usepackage{amsmath,amssymb}
\usepackage{enumerate}

\def \be{\begin{equation}}
\def \ee{\end{equation}}
\def \ba{\begin{array}}
\def \ea{\end{array}}
\def \beq{\begin{eqnarray}}
\def \eeq{\end{eqnarray}}

\begin{document}
\title{Microscopic diagonal entropy and its connection to basic thermodynamic relations}

\author{Anatoli Polkovnikov}
\affiliation{Department of Physics, Boston University, Boston, MA 02215}

\begin{abstract}

We define a diagonal entropy (d-entropy) for an arbitrary Hamiltonian system as $S_d~=~-\sum_n \rho_{nn}\ln \rho_{nn}$ with the sum taken over the basis of instantaneous energy states. In equilibrium this entropy coincides with the conventional von Neumann entropy $S_n=-{\rm Tr}\, \rho\ln\rho$. However, in contrast to $S_n$, the d-entropy is not conserved in time in closed Hamiltonian systems. If the system is initially in stationary state then in accord with the second law of thermodynamics the d-entropy can only increase or stay the same. We also show that the d-entropy can be expressed through the energy distribution function and thus it is measurable, at least in principle. Under very generic assumptions of the locality of the Hamiltonian and non-integrability the d-entropy becomes a unique function of the average energy in large systems and automatically satisfies the fundamental thermodynamic relation. This relation reduces to the first law of thermodynamics for quasi-static processes. The d-entropy is also automatically conserved for adiabatic processes. We illustrate our results with explicit examples and show that $S_d$ behaves consistently with expectations from thermodynamics.

\end{abstract}
\maketitle

\section{Introduction}

Explicit relations between thermodynamical and microscopical quantities and derivation of the laws of thermodynamics from microscopic theory have attracted attention of physics community for a long time~\cite{balian}. One of the main challenges is finding the correct microscopic definition of the entropy. While it is established that the von Neumann entropy $S_n=-{\rm Tr} \rho\ln\rho$ accurately describes equilibrium statistical ensembles, it clearly disagrees with the second law of thermodynamics because for isolated systems it is conserved in time for any processes~\cite{LL5}. This entropy also does not satisfy the fundamental thermodynamic relation, which relies on uniqueness of the entropy as a function of energy,  unless one makes an additional assumption that the density matrix always remains diagonal. It is well understood that the thermodynamic entropy of a closed system obtained from the von Neumann entropy,,which can also serve as a measure of information in the system, by coarse-graining increases with time (see e.g. Refs.~\cite{penrose, balian}). If properly defined coarse-graining can lead to correct predictions in complex systems subject to dynamical processes (see e.g. Refs.~\cite{joel, joel1}). Also it can be shown that the von Neumann's entropy behavior is consistent with the second law of thermodynamics in open systems~\cite{gogolin_thesis}. Yet the whole situation that on top of the microscopic description one needs to introduce a non-uniquely defined coarse-graining procedure, which is not part of the Hamiltonian dynamics, is not satisfactory.

Microscopic description of thermodynamics using the von Neumann entropy is problematic on several other reasons. For example, consider a sufficiently complex system that was subject to a process which started and ended in a distant past, and eventually achieved some steady state. By the ergodic hypothesis time average of any thermodynamic observable should be equivalent to the equilibrium ensemble average. For any observable $\Omega$, assuming that the spectrum is non-degenerate or that degeneracies are not important, its time average can be written as $\overline{\Omega}=\sum_n \Omega_{nn}\rho_{nn}$, where $\rho_{nn}$ are the time-independent diagonal elements of the density matrix in the eigenbasis of the Hamiltonian and $\Omega_{nn}$ are the diagonal matrix elements of the operator $\Omega$. All information about arbitrary time-averaged observables and thus about the steady state is contained in diagonal elements of $\rho$ (see Refs.~\cite{rigol_08, reimann} for more discussion). At the same time the von Neumann entropy explicitly depends on the nonlinear combination of the off-diagonal elements of the density matrix, which do not average to zero. So we have the situation that (i) the von Neumann entropy contains additional information, which does not appear in any thermodynamic measurement and (ii) its time average is different from the entropy of the equilibrium ensemble.

There is another important indication that the entropy should be sensitive only to the diagonal matrix elements of $\rho(t)$. It follows from the basic thermodynamics that in adiabatic processes the entropy of the system does not change~\cite{LL5}. At the same time the slow changes in the Hamiltonian do not induce transitions between instantaneous energy levels~\cite{LL3}, which implies that diagonal elements of the density matrix (and only they) do not change in time. We note that it is practically impossible to completely avoid transitions between energy levels in macroscopic systems~\cite{balian, np}. The proper adiabatic limit, therefore, should be defined as such when the heat (or excess energy) generated during a dynamical process is small. In turn heating of the system is related to the transitions between different instantaneous energy levels~\cite{np,ap_heating}. The thermodynamic adiabatic limit emerges as a result of shrinking phase space available for the transitions between instantaneous energy levels as the process becomes slower. Thus the heating is again sensitive only to the diagonal matrix elements of $\rho(t)$. Moreover, one can rigorously prove that for unitary evolution this heating is non-negative for any cyclic process (and under very generic assumptions for noncyclic processes) as long as the initial density matrix satisfies the conditions of passivity: it is stationary (diagonal) and monotonically decreasing function of energy~\cite{thirring, armen_kelv, armen_mwp, ap_heating, boksenbojm}. This statement actually coincides with the second law of thermodynamics in the Kelvin's form~\cite{kardar}. If the second law of thermodynamics is established from microscopics in one formulation, it should also follow from microscopics in other formulations involving entropy.

\section{Definition and properties of diagonal entropy.}

\subsection{Definition}

The considerations outlined in the previous section suggest a simple resolution. Let us define the following diagonal entropy (or simply d-entropy):
\be
S_d=-\sum_n \rho_{nn}\ln\rho_{nn}.
\label{ent_def}
\ee
There is some ambiguity in this definition associated with possible degeneracies in the energy spectrum. However, these degeneracies are usually either accidental or related to additional symmetries of the system. Generically these generacies are absent and we will assume here that this is indeed the case. If the Hamiltonian is time independent then $S_d$ formally corresponds to the usual von Neumann entropy defined by the time-averaged density matrix
\be
\overline \rho=\lim_{t\to\infty}1/t\int_0^t \rho(t')dt'
\ee
(see e.g. Ref.~\cite{penrose}). At the same time, the expression~(\ref{ent_def}) is defined instantaneously and thus does not require coarse-graining of the density matrix or the assumption that the Hamiltonian is stationary. The definition~(\ref{ent_def}) suggests a classical generalization where instead of a sum over many-body states one takes the integral over orbits corresponding to different energies of the system. The analogue of $\rho_{nn}$ would be a probability to find the system in the corresponding orbit:
\be
p(E)=\int d{\bf X}d{\bf P} \, \rho({\bf X},{\bf P})\delta(E-E({\bf X},{\bf P})),
\ee
where ${\bf X}={\bf x_1,x_2,\dots}, {\bf P={\bf p_1,p_2,\dots}}$ span the many-particle phase space and $\rho({\bf X}, {\bf P})$ is the probability to occupy a particular phase-space point. So the classical diagonal entropy can be defined as
\be
S_d^{\rm cl}=-\int dE N(E) p(E)\ln p(E),
\ee
where
\be
N(E)=\int d{\bf X}d{\bf P}\, \delta(E-E({\bf X},{\bf P}))
\ee
is the (many-particle) density of states.

\subsection{Diagonal entropy and the second law of thermodynamics}
\label{sec:second law}

Let us show that $S_d$ is consistent with the known properties of the thermodynamic entropy. In equilibrium, when the density matrix is stationary (diagonal), clearly $S_d=S_n$. Thus $S_d$ satisfies such requirements as extensivity, positivity. It also automatically vanishes in the zero-temperature limit satisfying the third law of thermodynamics. Then we observe that the entropy $S_d$ can change in time. If the initial state is stationary then for any time-dependent process in a closed Hamiltonian system we have $S_d(t)\geq S_d(0)$.  The proof of this statement can be adopted from Ref.~\cite{penrose} (see Eqs.~(2.4) - (2.10)) by straightforward generalization to the quantum case where instead of the time averaging we use the microscopic definition of $S_d(t)$. First, we identify $\rho^d(t)$ as a
diagonal part of the full time-dependent density matrix $\rho(t)$, i.e. $\rho^d_{nn}(t)=\rho_{nn}(t)$ and $\rho^d_{nm}(t)=0$ for $n\neq m$. By using the Jensen's inequality we obtain that (see also Eq.~(1.40) in
Ref.~\cite{wehrl})
\be
{\rm Tr} (-\rho\ln\rho+\rho_d\ln\rho_d)\leq {\rm Tr} (\rho-\rho_d)\ln\rho_d=0
\ee
Hence we find that $S_d(t)\geq S_n(t)$. At the same time in closed systems the von Neumann entropy is conserved in time, $S_n(t)=S_n(0)$. Noting that $S_d(0)=S_n(0)$, since the initial density matrix is diagonal by the assumption, we prove
\be
S_d(t)\geq S_d(0).
\label{second_law}
\ee
Note that Eq.~(\ref{second_law}) does not imply that $S_d(t)$ is necessarily a monotonic function of time.

There is a simple physical reason behind the diagonal entropy increase. For unitary evolution the density matrix at time t is given by
\be
\rho(t)=U^\dagger \rho(0) U,
\ee
where $U$ is the evolution operator. If the initial density matrix is stationary (diagonal) then it is easy to see that the diagonal elements of the time evolved density matrix are
\be
\rho_{nn}(t)=\sum_m |U_{mn}|^2 \rho_{mm}(0)=\sum_m P_{nm}\rho_{mm}(0).
\ee
The matrix elements $P_{nm}=|U_{mn}|^2$ have a simple interpretation of transition rates between instantaneous energy levels. The matrix $P$ is doubly stochastic implying that the transition probabilities satisfy the sum rule~\cite{ap_heating}:
\be
\sum_m P_{mn}=\sum_m P_{nm}=1.
\ee
This sum rule states that the total rate of leaving a particular level $m$ is equal to the sum of rates of coming to this level from all other energy levels. This sum rule is a direct consequence of the unitarity of the evolution and it is not generally valid in open (not-Hamiltonian) systems. It is straightforward to realize that the sum rule above guarantees that any dynamical process leads to a more uniform spread of the probability density among energy levels, i.e. forces system closer to a microcanonical distribution $\rho_{nn}=const$. Since the diagonal entropy is a measure of the spread of $\rho_{nn}$ it necessarily increases or stays constant.

\subsection{Diagonal entropy and the fundamental thermodynamic relation.}

One of the most important and nontrivial postulates of thermodynamics is that the entropy is a unique function of energy and external parameters: $S=S(E,\lambda_1,\lambda_2,\dots)$. This assumption allows to relate infinitesimal change of entropy for {\em any} dynamical process to the change of energy and change of external parameters
\be
\Delta E=T\Delta S-\sum_j {\partial E\over\partial \lambda_j}\biggr|_S\Delta\lambda_j.
\label{fund_rel1}
\ee
The derivatives of energy with respect to the external parameters $\lambda_j$ are called generalized forces. They are related to adiabatic changes of the spectrum of the system~\cite{balian}. Therefore the first term $T\Delta S$ (heat) must be associated with the transitions between microscopic energy levels. The non-triviality of Eq.~(\ref{fund_rel1}) is that it must be valid for any microscopic process as long as changes of energy, entropy, and parameters of the system remain small. In particular, it implies that for any cyclic process the change of energy is simply proportional to the change of entropy with the proportionality constant (temperature) being completely insensitive to how and which external parameters changed in time. This thermodynamic postulate thus puts very severe constraints on the microscopic definition of the entropy. Von Neumann entropy clearly does not satisfy Eq.~(\ref{fund_rel1}). For example, if we consider any cyclic process in a closed system then $\Delta S_{\rm n}\equiv 0$, while $\Delta E$ is generically nonzero.

Let us show that the d-entropy satisfies the fundamental relation (\ref{fund_rel1}). For now we will do it making an additional assumption that the initial density matrix is thermal, i.e. that it is described the Gibbs distribution:
\be
\rho^0={1\over Z}\exp(-\beta H^0).
\label{gibbs_rho}
\ee
In the next section we will lift this assumption and will show that Eq.~(\ref{fund_rel1}) is valid in a more general case.

 First we write the change of the energy to the linear order in $\Delta\rho$ and $\Delta\lambda_j$:
\be
\Delta E\approx \sum_n \Delta E_n(t) \rho_{n}^0 +\sum_n E_n(0)\Delta \rho_{nn}(t),
\label{delta_E}
\ee
where $\Delta E_n(t)=E_n(t)-E_n(0)$ is a change of the instantaneous energy levels due to time evolution, $\Delta\rho_{nn}(t)$ is the change of the diagonal matrix elements of the density matrix, and $\rho_n^0\equiv \rho_{nn}(0)$. In the adiabatic limit $\Delta \rho_{nn}(t)=0$ and thus the first term in Eq.~(\ref{delta_E}) corresponds to the adiabatic change of the energy $\Delta E_{\rm ad}(t)$ while the second one corresponds to the heat~\cite{ap_heating}. Next we consider a similar expression for the change of the d-entropy. To the leading order in $\Delta \rho_{nn}(t)$ (noting that $\sum_n\Delta\rho_{nn}(t)=0$) we find
\be
\Delta S_d\approx-\sum_n \Delta\rho_{nn}(t)\ln\rho_{n}^0.
\label{delta_S}
\ee
Using the Gibbs distribution~(\ref{gibbs_rho}) and comparing Eqs.~(\ref{delta_E}) and (\ref{delta_S})
we immediately find
\be
\Delta E\approx \Delta E_{\rm ad}+ T\Delta S_d.
\label{delta_E1}
\ee
 The first term here $\Delta E_{\rm ad}$ is a function of the state, i.e. it depends only on the instantaneous values of the external parameters $\lambda_j$ and the initial probabilities $\rho_n^0$. Thus it can be expressed
 \be
 \Delta E_{\rm ad}=\sum_j (\partial E/\partial\lambda_j)_{S_d}\Delta \lambda_j.
 \ee
We see that Eq.~(\ref{delta_E1}) is indeed equivalent to the relation~(\ref{fund_rel1}). Note that actually the derivation of Eq.~(\ref{delta_E1}) relied neither on the assumption that the off-diagonal elements of the diagonal density matrix are zero nor on the assumption that the system was closed during the dynamical process.

\subsection{Diagonal entropy and the energy distribution. Equivalence of ensembles from the density matrix.}

It is well known from standard thermodynamics that in large systems the choice of the statistical ensemble is usually dictated purely by convenience. The basic laws of thermodynamics, in particular the fundamental relation~(\ref{fund_rel1}), must be valid for any ensemble. To see such equivalence from microscopics and to see which requirements on microscopic ensembles are necessary to recover Eq.~(\ref{fund_rel1}) from Eq.~(\ref{delta_E1}) let us perform some simple manipulations with diagonal entropy:
\beq
\label{sd_en}
S_d&=&-\sum_n \rho_{nn}\ln(\rho_{nn})\nonumber\\
&=&-\int dE\, N(E) \rho^d(E)\ln\left[{\rho^d(E)N(E)\delta E\over N(E)\delta E}\right]\\
&=&\int dE\, W(E)S_{\rm micro}(E)-\int dE\, W(E)\ln\left[W(E)\delta E\right].\nonumber
\eeq
Here
\be
N(E)=\sum_n \delta(E-E_n)
\ee
is the many-body density of states,
\be
\rho^d(E)=\sum_n \rho_{nn}\delta(E-E_n),
\ee
$W(E)=\rho^d(E) N(E)$ is the distribution function of energy. It satisfies the normalization condition $\int dE\, W(E)=1$ and defines different moments of energy, e.g. $\langle E\rangle=\int dE\,E\,W(E)$. And finally
\be
S_{\rm micro}(E)=\ln \left[N(E) \delta E\right]
\ee
is the microcanonical entropy and $\delta E=\sqrt{\langle E^2\rangle -\langle E\rangle^2}$ is the width of the distribution. Note that the factor $\delta E$ here is arbitrary, only needed to fix the dimensions. In macroscopic systems $N(E)$ is exponentially large, while $\delta E$ is not so the contribution from this factor is negligible~\cite{kardar}.

Let us now look closely to Eq.~(\ref{sd_en}). The first term reduces to usual thermodynamic entropy as long as the distribution $W(E)$ is sufficiently narrow on the scale of change of the equilibrium entropy, which is typically some extensive scale. So as long as $W(E)$ is sufficiently narrow in energy such that $\delta E$ is subextensive the first term in Eq.~(\ref{sd_en}) just reproduces the conventional thermodynamic entropy plus non extensive corrections. In non-equilibrium systems it can be shown that the distribution $W(E)$ is indeed narrow under very general conditions such as the locality of the Hamiltonian~\cite{biroli09}.

Thus we see that in order to satisfy the equivalence between a given ensemble defined by the density matrix $\rho$ and equilibrium microcanonical ensemble the first necessary requirement is that energy fluctuations are subextensive.
The second term in Eq.~(\ref{sd_en}) is more subtle. If we assume that $W(E)$ is a smooth function of energy then it is clear that this term is always subextensive. However, in true non-equilibrium ensembles there can be large state to state fluctuations between occupancies of different microscopic energy levels (see e.g. Ref.~\cite{roux}) thus we can write $W(E)=\tilde W(E)\,\sigma(E)$ where $\tilde W(E)$ is a smooth part of the distribution $W(E)$ and $\sigma(E)$ is a stochastic part. Note that low moments of the energy are determined by $\tilde W(E)$ therefore only $\tilde W(E)$ is measurable. In noninegrable systems according to the Berry's conjecture~\cite{berry} and its generalizations the many-body eigenstates can be considered as pseudo-random vectors in the Hilbert space. Thus one can expect that $\sigma(E)$ is a random Gaussian function. This is indeed consistent with recent numerical results~\cite{roux, rigol_09, clemens}. Such Gaussian noise clearly contributes a subextensive correction to the first term in Eq.~(\ref{sd_en}) and can be neglected. The situation can change in integrable systems, where due to the dynamical constraints only an exponentially small subset of many-body states can be occupied~\cite{clemens}. In other words the density matrix $\rho$ is very (exponentially) sparse. Then the contribution from $\sigma(E)$ to the diagonal entropy is extensive and the latter is different from the equilibrium temperature. The sparseness of the distribution is also necessary to see deviations of expectation values of other observables from equilibrium values~\cite{biroli09}. Therefore in this sense the diagonal entropy is not different from the other observables.

To summarize the discussion above the diagonal entropy in macroscopic systems is equivalent to equilibrium microcanonical entropy up to subextensive corrections if

\begin{enumerate}[i]
\item the energy distribution $W(E)$ is narrow so that $\delta E$ is subextensive,
\item the energy distribution (or the diagonal part of the density matrix) is not very (exponentially) sparse so that $\int dE\,W(E)\ln(W(E)/\delta E)$ is subextensive.
\end{enumerate}

Both assumptions are expected to be valid in non-integrable systems subject to arbitrary dynamical process. This was indeed verified in a number of recent studies~\cite{biroli09, clemens}. Under the second condition one can simplify Eq.~(\ref{sd_en}) ignoring the stochastic term and substituting $W(E)$ by its smooth part $\tilde W(E)$ so that
\be
S_d\approx \int dE\, \tilde W(E)S_{\rm micro}(E)-\int dE\, \tilde W(E)\ln(\tilde W(E)\delta E).
\label{Sd_approx}
\ee
This equality also shows that under these assumptions the diagonal entropy is measurable since $\tilde W(E)$ can be extracted from measuring energy distribution. Note that $S_{\rm micro}$ is defined through the many-body density of states, which does not depend on the dynamical process. It is easy to see that if the conditions (i) and (ii) are satisfied for the initial state then the diagonal entropy still satisfies the fundamental relation~(\ref{fund_rel1}). Moreover if these two conditions are satisfied at each moment in time, which should be the case for an arbitrary dynamical process in non-integrable systems, then the Eq.~(\ref{fund_rel1}) is valid at each moment in time and can be integrated. In Sec.~\ref{sec:examples} we will illustrate this numerically using a particular example.

\subsection{Additivity of diagonal entropy. Total entropy versus sum of local entropies.}

One of the important properties of the thermodynamic entropy in equilibrium is its additivity. For equilibrium ensembles the additivity of the statistical entropy follows e.g. from approximate multiplicativity of the number of microstates in two weakly coupled subsystems and the postulate of thermodynamics that all microstates are equally probable~\cite{reif}. Indeed according to the standard arguments the number of microstates in the system at a fixed energy $\Omega(E)$ is approximately equal to the product of number of microstates in two subsystems comprising the system: $\Omega(E)\approx \Omega_1(E_1^\star)\Omega_2(E-E_1^\star)$, where $E_1^\star,\,E-E_1^\star$ is the equilibrium partitioning of the energy between the two subsystems. The additivity of thermodynamic entropy immediately follows from the definition $S(E)=\ln\Omega(E)$.

The situation becomes much more subtle away from equilibrium. Some textbooks state that the non-equilibrium entropy is not well defined. Instead one needs to split the system into sufficiently small but yet macroscopic subsystems such that each subsystem is in approximate equilibrium but the subsystems are not in equilibrium with each other (see e.g. Ref.~\cite{LL5}). Other textbooks simply avoid discussing this issue. Within the first approach the notion of total entropy of the system looses its separate meaning like in equilibrium because it is not defined separately from the sum of the parts. On the contrary the entropy defined as $S(E)=\ln\Omega(E)$ is clearly not additive away from equilibrium instead it satisfies $S>S_1+S_2$. If the system is allowed to equilibrate then according to thermodynamics the sum of the entropies $S_1+S_2$ should increase and reach the equilibrium value $S$ where the additivity of the entropy is restored. Note that the total entropy of the system $S(E)=\ln \Omega(E)$ can change only if either the total energy of the system or external parameters change in time. The difference $S-S_1-S_2$ can serve as a measure of ``non-equilibriumness'' of the system.

Since thermodynamics gives specific predictions about entropy only in equilibrium, its precise definition away from equilibrium, from the point of view of thermodynamics, is not unique. Indeed e.g. the additivity or non-additivity of the entropy can only be established by comparing the total entropy to the sum of the parts. Away from equilibrium only sum of the parts, each in approximate thermodynamic equilibrium, can be measured from e.g. specific heat. Microscopically the situation that one needs to tie the definition of the object (entropy) to the particular state of the system is very inconvenient. It is much easier to keep the definition (\ref{ent_def}), which is equivalent to Eqs.~(\ref{sd_en}), (\ref{Sd_approx}), intact but keep in mind that the additivity of the d-entropy is expected only in equilibrium (or more accurately in steady) state. Thus we have to keep in mind the following correspondence
\beq
S_d\quad &\Leftrightarrow & \quad\ln\Omega(E)\\
S_d^{(1)}+S_d^{(2)}\quad &\Leftrightarrow &\quad \ln\Omega_1(E_1)+\ln\Omega_2(E_2),
\eeq
where $S_d^{(1)}$ and $S_d^{(2)}$ are the diagonal entropies of the two subsystems. In this way one avoids any ambiguities with the definition of the total entropy. But one needs to keep in mind that away from equilibrium sum of thermodynamics entropies of subsystems corresponds to the sum of diagonal entropies of these subsystems but not to the total d-entropy.

In thermal equilibrium the additivity of the d-entropy follows from the additivity of von Neuman entropy, which is well established in statistical physics. If a non-integrable system is driven away from equilibrium and then allowed to equilibrate then the additivity of the d-entropy in the new steady state follows from Eq.~(\ref{Sd_approx}) under two conditions: (i) the equilibrium microcanonical entropy is approximately additive and (ii) the distribution of the energy partitioning between the two systems is narrow. The former property is automatically satisfied in macroscopic systems while the latter is expected from extensivity of the energy. In the next section we will illustrate how this extensivity approximately works using a numerical example. Note that for von Neumann's entropy one does not expect additivity to hold even after the system reaches the steady state.

Let us point another property of the sum of diagonal entropies, which is also in line with thermodynamic expectations. Assume that we have two initially uncoupled systems each in a local equilibrium. Then the two systems are allowed to interact. According to the second law of thermodynamics the sum of entropies of the two subsystems should either increase or stay the same. The diagonal entropy satisfies the same property~\cite{gogolin}:
\be
S_d^{(1)}(t)+S_d^{(2)}(t)\geq S_d^{(1)}(0)+S_d^{(2)}(0).
\label{second_law_sum}
\ee
The proof of this statement is as follows. Since the two systems are initially in local equilibrium then $S_d^{(1)}(0)=S_{\rm n}^{(1)}(0)$ and $S_d^{(2)}(0)=S_{\rm n}^{(2)}(0)$. Because the two systems are initially decoupled the von Neumann's entropy is additive $S_{\rm n}(0)=S_{\rm n}^{(1)}(0)+S_{\rm n}^{(2)}(0)$. The von Neumann's entropy is conserved in time thus $S_{\rm n}(t)=S_{\rm n}(0)$. Next we can use subadditivity of the von Neumann's entropy~\cite{lieb}: $S_{\rm n}(t)\leq S^{(1)}_{\rm n}(t)+S^{(2)}_{\rm n}(t)$. Finally using that $S_{\rm n}(t)\leq S_d(t)$ (see Sec.~\ref{sec:second law}) we prove Eq.~(\ref{second_law_sum}).

\section{Examples}
\label{sec:examples}

\subsection{Noninteracting particles in a box.}
\label{sec:box}

\begin{figure}[ht]
\includegraphics[width=3.5in]{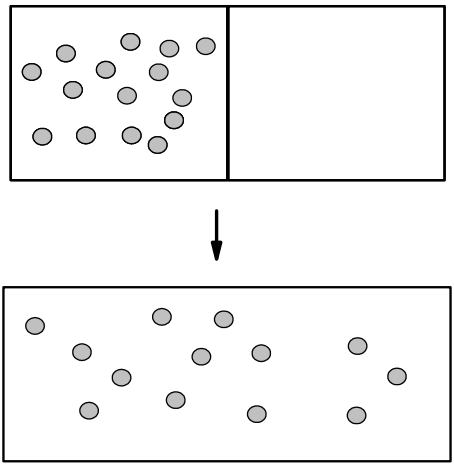}
\caption{Illustration to the example of expanding gas.}
\label{fig_gas}
\end{figure}

Let us start from a very simple but illustrative example highlighting qualitatively behavior of the diagonal entropy.
Namely we consider a classical noninteracting gas initially confined to the one half of a container separated by a membrane (see Fig.~\ref{fig_gas}). At moment $t=0$ the membrane is removed and the gas expands to the whole container. From the point of view of classical thermodynamics the entropy of the gas increases by the amount $\Delta S=N\ln 2$, where $N$ is the total number of particles. It is easy to see that the same result applies to the d-entropy. One can obtain it by explicit calculations re-expanding the initial density matrix in the new basis and computing d-entropy. However, there is a simpler way to see this. As we double the volume the density of the momentum states per particle doubles, because the momentum is quantized in units of $2\pi/(L/2)$ before the expansion and $2\pi/L$ after the membrane is removed. This means the the same momentum distribution (momenta of particles do not change in this process) is now projected to the twice the number of energy states so that by conservation of probability diagonal elements of the density matrix for each particle are reduced by a factor of two $\rho_{nn}\to {1\over 2}\rho_{nn}$. Thus we immediately see that d-entropy per particle increases by $\ln(2)$ so that $\Delta S_d=N\ln(2)$. This example also highlights the issue of non-additivity of the d-entropy away from equilibrium (or more generally steady state). Indeed the jump of $S_d$ occurs immediately after the membrane is removed. At the same time the local d-entropies corresponding to the left and right parts of the container do not change right away. We expect that in the chaotic cavity the sum of $S_{1d}$ and $S_{2d}$ will gradually increase in time and after the gas reaches the steady state this sum will approach $S_d$. Then at intermediate times the difference between $S_d$ and $S_{1d}+S_{2d}$ is a possible measure characterizing how far the system is away from reaching the steady state. This example also highlights importance of quantum mechanics even in this purely classical problem. Note that the d-entropy changes here purely because we double the number of microscopic single particle quantum states.

\subsection{Hard core bosons in two dimensions.}

\begin{figure}[ht]
\includegraphics[width=8cm]{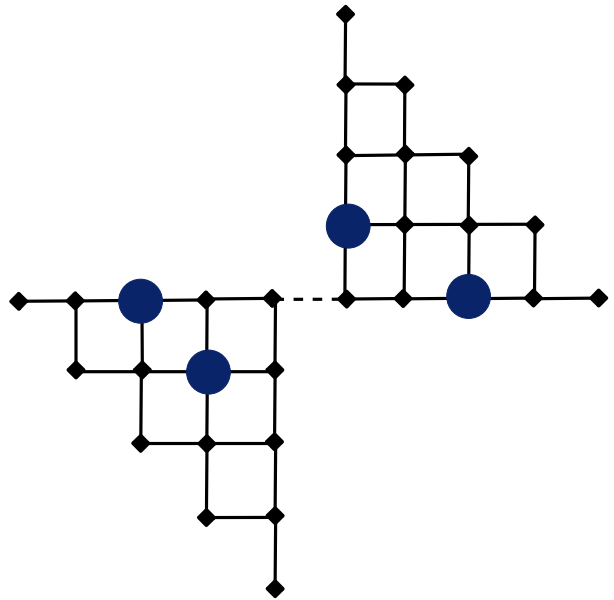}
\caption{Lattice system corresponding to the second example. The system represents four hard core bosons moving in two quantum dots coupled by a weak link characterized by the hopping amplitude $J_1$ (dashed line). All other links have unit hopping $J=1$.}
\label{fig:hard_core}
\end{figure}

The next example we will analyze involves dynamics of hard core bosons in two quantum dots connected by a weak link. The system we will analyze is similar to that used in Ref.~\cite{rigol_08} (see Fig.~\ref{fig:hard_core}). This is a non-integrable closed system which shows clear signatures of thermalization.
In this section we will systematically analyze different properties of d-entropy in this system for various dynamical protocols.

\subsubsection{Local and global d-entropies after a quench.}

First we will analyze an example similar to that in Sec.~\ref{sec:box}. Namely, we will consider a system of four bosons in initially decoupled dots, $J_1=0$ (see Fig.~\ref{fig:hard_core}). We will consider two initial setups. In the first setup each dot initially contains exactly two particles in the ground state. In the second setup all four particles are prepared in the ground state of the left dot. Then suddenly at time $t=0$ the coupling connecting two dots is turned on to a particular value $J_1=0.5$.

\begin{figure}[ht]
\includegraphics[width=9cm]{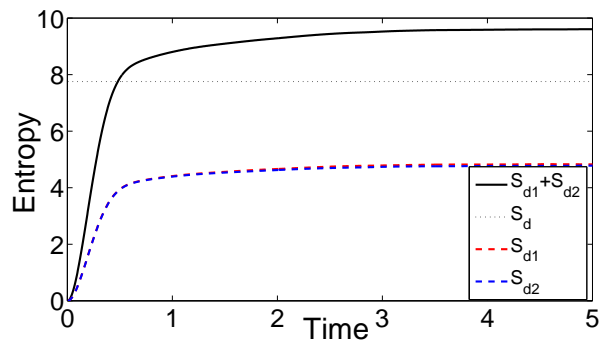}
\caption{Diagonal entropy as a function of time after the tunneling quench. The initial state corresponds to
two decoupled dots (see Fig.~\ref{fig:hard_core}) with two particles in each prepared in the ground states of each dot. The thin dashed line shows the diagonal entropy of the system, it is constant after the quench. Two thick dashed lines represent diagonal entropies of each individual dot. The thick black line is their sum. }
\label{fig:diag_ent_1}
\end{figure}

On Fig.~\ref{fig:diag_ent_1} we plot time dependence of the full diagonal entropy as well as of the local diagonal entropies and their sum. The full diagonal entropy is constant after the quench because the probabilities of occupying energy eigenstates $\rho_{nn}$ do not change in time if the Hamiltonian is constant. At the same time the local diagonal entropies of the subsystems increase with time and saturates at equilibrium values. The sum of local entropies $S_{d1}+S_{d2}$ also saturates at the value somewhat bigger than $S_d$. This overshooting is expected even in equilibrium because of the finite size effects. Indeed it is easy to see that $\log{D}\approx 8.49<\log(D_1*D_2)\approx 11.91$, where $D$ is the Hilbert space size of the total system with exactly four particles and $D_1=D_2$ are the Hilbert space sizes of the two subsystems (dots) with up to four particles in each. Only in the thermodynamic limit when the number of particles becomes macroscopic the full thermodynamic entropy becomes equal to the sum of the entropies of the two subsystems. We expect the same to be true for the diagonal entropy.

\begin{figure}[ht]
\includegraphics[width=9cm]{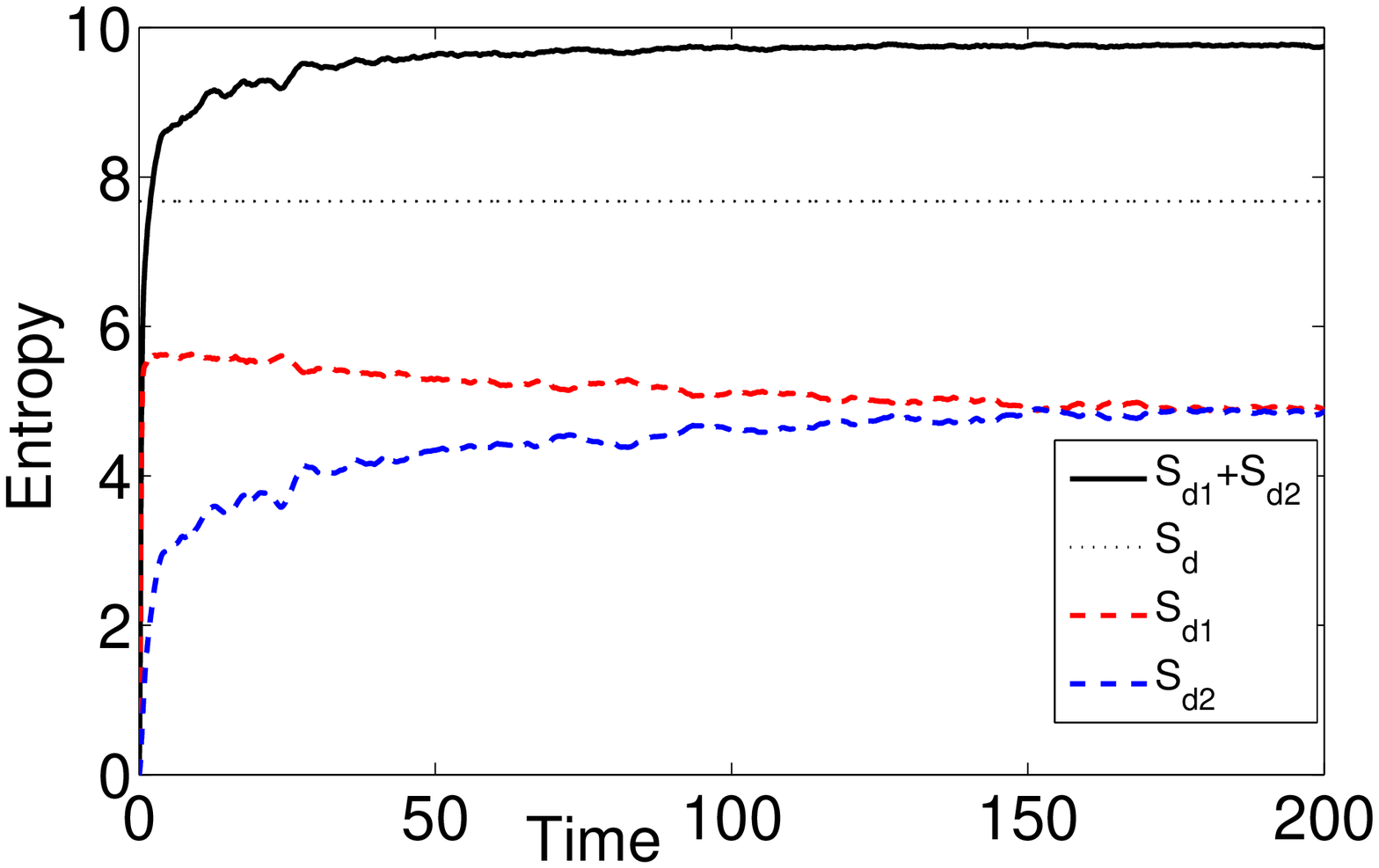}
\includegraphics[width=9cm]{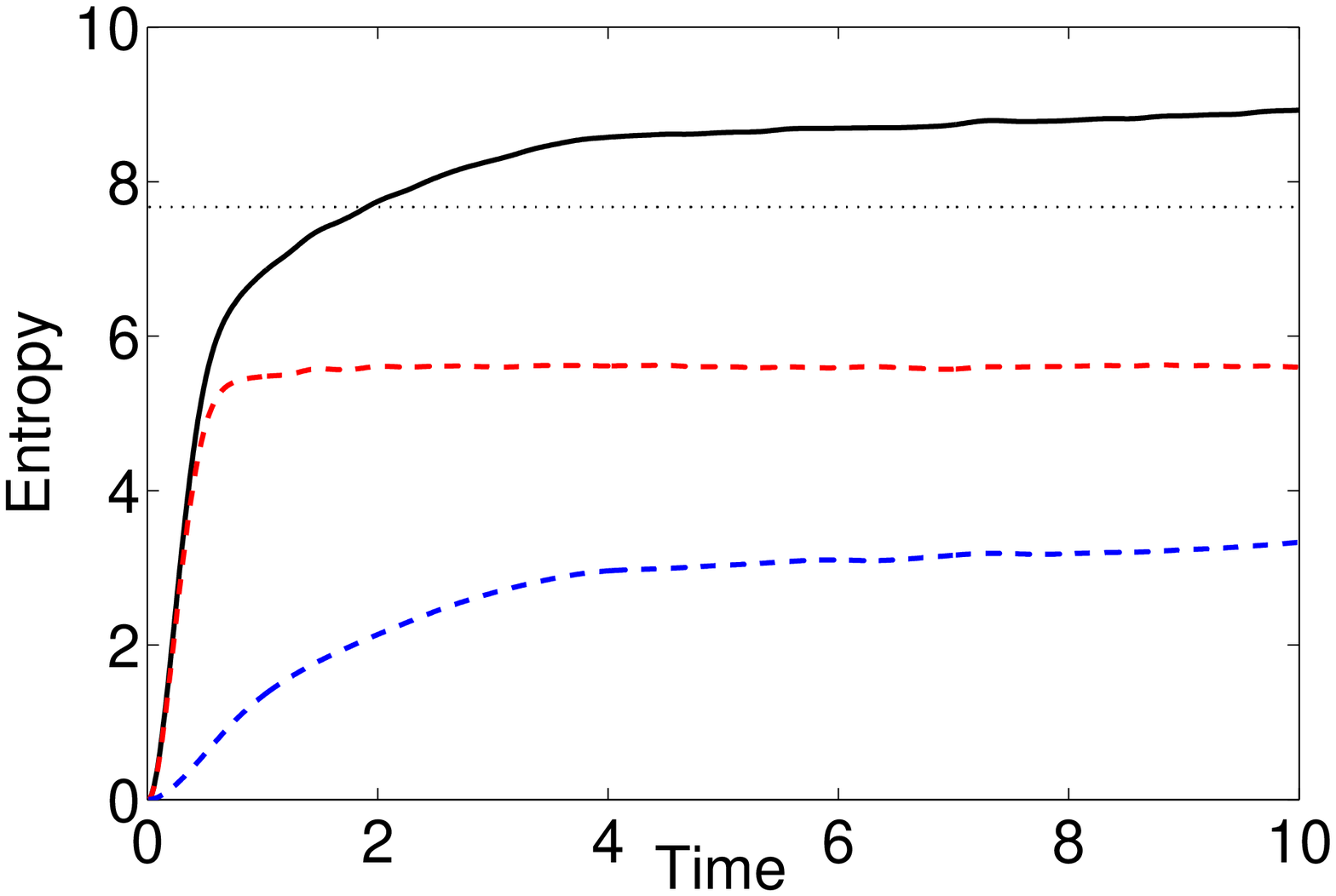}
\caption{Same as in Fig.~\ref{fig:diag_ent_1} except that initially all four particles are prepared in the ground state of the left dot. The bottom graph shows short time dynamics of the entropy.}
\label{fig:diag_ent_2}
\end{figure}

On Fig.~\ref{fig:diag_ent_2} we plot similar time dependence of the entropies for the second initial condition, where all particles are initially placed into the ground state of the left dot (subsystem 1). There the dynamics is somewhat richer. At short times (see the bottom graph) the first (left) dot rapidly thermalizes due to opening the link and changing the structure of eigenstates. The entropy of the second dot increases at much longer time scales due to slow tunneling of particles. During the same long time scales the diagonal entropy of the first dot somewhat decreases due to the fact that it looses particles. The sum of local entropies however increases as it should. Eventually the system comes to the equilibrium. One can check that at this point the average number of particles becomes the same in both subsystems ($N_1=N_2=2$). The sum of the diagonal entropies of the two dots again somewhat overshoots the total diagonal entropy due to the finite size effects.

\subsubsection{Diagonal entropy and time reversibility.}

A very important issue in understanding emergence of laws of thermodynamics from microscopic Hamiltonian dynamics is the issue of time reversability. In the general discussion we argued that this contradiction is very illusive: in order to prepare the system in the low entropy state it is necessary to open it, e.g. connecting it to a refrigerator. Dynamics of the open system is necessarily non-Hamiltonian (although dynamics of the system plus the bath is) so there is no microscopic reason why the entropy can not be lowered. If one closes the system and considers an arbitrary dynamical processes within the Hamiltonian dynamics then as we proved the diagonal entropy will increase (or at most stay constant e.g. in the adiabatic limit) as long as the initial state is stationary. In principle it is possible to lower the entropy back to the original value by doing exact time-reversal transformation. However, this time-reversal transformation should be done with a very high precision, otherwise with probability exponentially close to unity the entropy will continue increasing (see Ref.~\cite{joel} for related discussions). While we can not give a general mathematical proof that this is always the case, we will present here the numerical evidence for such a scenario.

We will again analyze the system plotted in Fig.~\ref{fig:hard_core} but now for computational reasons with 18 sites in total (with two top left sites missing in the right dot) but still four particles. We will consider a process where we repeatedly change the coupling $J_1$ connecting the two dots between the two values $J_1=0.5$ and $J_2=1$ starting from the ground state with the coupling $J_1$. Between each quench of $J$ we will wait for a time $T$ randomly distributed in the interval $[0,1000]$. Such randomness is introduced to avoid effects of revivals characteristic for finite size systems driven by periodic modulations. On average the jumps between quenches of $J$ are longer than the relaxation times in the system (see Figs.~\ref{fig:diag_ent_1}, \ref{fig:diag_ent_2}) so that we can think that each quench starts from a stationary state. On a coarse-grained time scales this sequence of quenches gives one of the simplest realizations of a quasi-static process in a closed system.

\begin{figure}[ht]
\includegraphics[width=9cm]{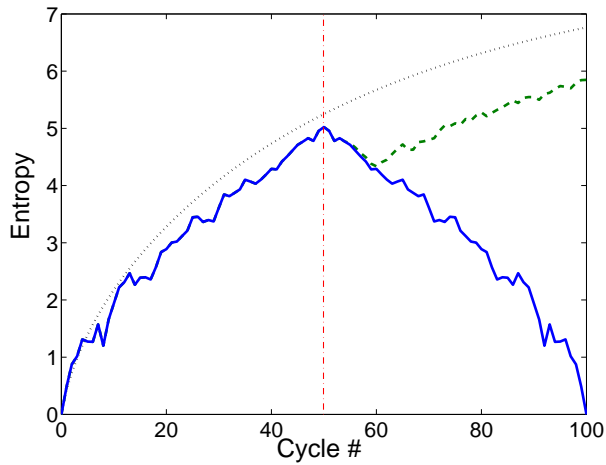}
\caption{Diagonal entropy as a function of number of cycles for a repeated quench protocols where in each cycle the coupling $J_1$ (see Fig.~\ref{fig:hard_core}) is quenched from $J_1=0.5$ to $J_1=1.0$ and then back. The waiting time between quenches is randomly distributed in the interval $[0,1000]$. After fifty cycles a time reversal transformation is applied to the system. Blue solid line corresponds to the exact evolution. Black dashed line describes coarse-grained dynamics where off-diagonal matrix elements are averaged to zero after each quench (this corresponds to time averaging density matrix between quenches). Green dashed line corresponds to exact dynamics where time reversal symmetry is weakly violating by introducing weak random uniform energy shift in the system uniformly distributed in the interval $[0,10^{-4}$]. }
\label{fig:diag_ent_3}
\end{figure}

We show the resulting evolution of the diagonal entropy after such a series of quenches in Fig.~\ref{fig:diag_ent_3}. The blue solid line describes the exact time evolution. The black dashed line describes coarse-grained dynamics where after each quench the density matrix is effectively averaged over time. This is equivalent of quenching the system each time from the diagonal ensemble obtained from the previous quench. It is clear from the figure that until cycle \# 50, where a time reversal transformation is introduced, the exact and coarse-grained dynamics give very similar time dependence. In both cases the d-entropy gradually approaches maximum possible entropy in the system $S_{\rm max}=\log D\approx 8.03$, where $D$ is the dimension of the Hilbert space. The situation changes drastically after we introduce the time reversal transformation. Then the exact dynamics results in gradual decrease of the d-entropy as the system is quenched back to the ground state, while in the coarse-grained dynamics is insensitive to the time reversal transformation and the d-entropy continues to increase. To check the sensitivity of the entropy decrease to the accuracy of the time-reversal transformation we also analyzed exact dynamics when we introduced small time-reversal symmetry perturbation (green dashed line). Namely after reversing the dynamics after each cycle we introduced a small random uniform energy shift in the system distributed in the interval $[0,10^{-4}]$. As can be seen from the figure, this small perturbation results in initial entropy decrease after time reversal, which persists for a few cycles and after that d-entropy starts increasing again. This is indeed in perfect agreement with our expectations.

\subsubsection{Uniqueness of the diagonal entropy as a function of the energy.}

Finally using the same numerical setup we will check the key property of the entropy - its uniqueness as a function of energy. For this purpose we will consider a non-equilibrium dynamics in the system where, as in the previous example, the coupling $J_1$ is quenched back and force. But now we will consider periodic sequence of pulses where the time between consequent quenches is fixed. This results in a highly non-equilibrium dynamics non monotonic dependence of both entropy and energy on the number of cycles. This dependence is also very sensitive to the period of the modulation.

\begin{figure}[ht]
\includegraphics[width=9cm]{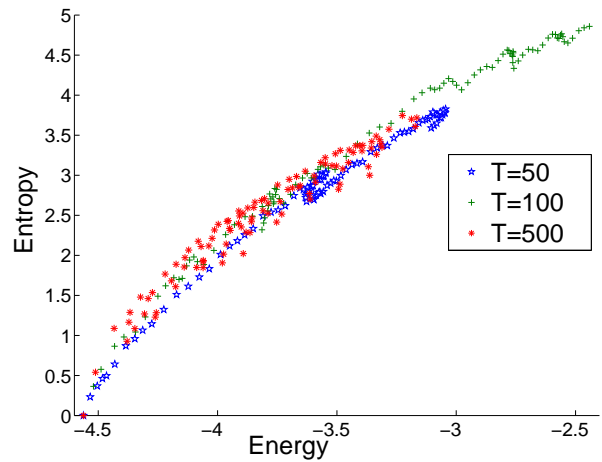}
\caption{Entropy versus energy for the periodic sequence of quenches of $J_1$. The setup is the same as in Fig.~\ref{fig:diag_ent_3} but the time between pulses $T$ is fixed. The three plots correspond to $T=50,\, 100,\, 500$ respectively.}
\label{fig:diag_ent_4}
\end{figure}

On Fig.~\ref{fig:diag_ent_4} we plot three different dependence of entropy versus energy for different times between quenches: $T=50,\,100,\,500$. It is easily seen that all three curves are very close to each other despite dependence of both energy and entropy on time strongly differ. Small discrepancies between the curves are expected because of relatively small system size. The should vanish as the system size increases. One can also consider pulses with different modulation amplitudes. The resulting curves again show only small deviations from the ones plotted in Fig.~\ref{fig:diag_ent_4}.

\section{Summary.}

We showed that the d-entropy $S_d$, introduced in this work, is consistent with various properties of the thermodynamic entropy. The d-entropy is microscopically defined. For equilibrium ensembles it coincides with the von Neumann's entropy. However, away from equilibrium, it is different. We showed that the d-entropy satisfies the key properties of the thermodynamic entropy:

\begin{itemize}
\item If a closed system is initially prepared in a stationary state then the change of the d-entropy as a result of {\em any} dynamical process is non-negative.

\item The d-entropy is automatically conserved for adiabatic processes characterized by vanishing transition probabilities between instantaneous energy eigenstates (or in macroscopic systems by vanishing phase-space volume available for such transitions).

\item If the two initially stationary subsystems brought to the contact with each other then the sum of d-entropies either increases or stays constant.

\item If initially the system is prepared in the thermally equilibrium state at temperature $T$ then for {\em any} dynamical process for both open and closed systems the d-entropy satisfies the fundamental thermodynamic relation: $dE=TdS_d-\sum_{\lambda_i}  {\partial E\over\partial\lambda_i}_{S_d}d\lambda_i$, where $\lambda_i$ are external parameters describing the system.

\item The d-entropy is uniquely related to the energy distribution and thus is measurable (at least in principle) both in equilibrium and away from it. Moreover as long as the Hamiltonian of the system is local and non-integrable, such that the energy fluctuations are subextensive and the support of the density matrix in the eigenbasis of the Hamiltonian is not exponentially sparse, the d-entropy is a unique function of the energy at any time (whether the system is in equilibrium or not). This means, in particular, that the fundamental thermodynamic relation is microscopically applicable to any dynamical processes. If the process is non-stationary then by temperature and generalized forces one needs to understand their equilibrium values.

\item The diagonal entropy is non-additive away from equilibrium. In particular, in macroscopic systems the difference between the total d-entropy and the sum of d-entropies of the subsystems can serve as a measure of non-equilibriumness of the system again in agreement with thermodynamics, Using a specific example of a non-integrable system we illustrated that after an instantaneous quench the total d-entropy changes instantaneously (similarly to the total energy) while the sum of the entropies of the subsystems changes gradually (again similarly to the sum of energies of the subsystems).

\end{itemize}

Let us make a few additional remarks. First we note that in complex closed systems for quasi-stationary processes the density matrix always remains effectively diagonal in a sense that off-diagonal matrix elements of density matrix having essentially random phases do not affect any local observables~\cite{rigol_08} and the consequent dynamics. In this case as we demonstrated $S_d$ should monotonically increase in time. In such situations one can effectively erase off-diagonal elements of $\rho$ and assume the density matrix to be always diagonal. Then the diagonal entropy simply coincides with the von Neumann entropy. However, in general this assumption is not satisfied, for example, if the process is not quasi-stationary or if the system has memory effects. Then erasing off-diagonal elements of the density matrix is not permitted. For example, in simple spin systems one can perform spin-echo type experiments, where after quenching magnetic field and initial dephasing in the system due to some randomness, one can perform a time-reversal transformation so that the spins restore the original coherence. In this case, the d-entropy first increases and then decreases back in time to the original value. But the situations like this where the entropy can decrease in time are very non-generic because they require ability to perform a time-reversal transformation on the Hamiltonian very accurately (see also discussion in Refs.~\cite{joel, boksenbojm}). In Fig.~\ref{fig:diag_ent_3} we illustrate this point with a toy-model system subject to the spin-echo type process. Even a very small perturbation, which breaks time inversion, completely destroys the reversibility and the d-entropy continues to increase in time. From our analysis it follows that the maximum entropy state corresponding to $\rho_{nn}$ independent of $n$ is a natural attractor of the Hamiltonian dynamics. In this respect, the second law of thermodynamics naturally follows from the microscopic equations of motion.

There is a more subtle issue of relevance of the d-entropy to information. If the Hamiltonian is constant in time, $S_d$ is only sensitive to the stationary information encoded in time-independent diagonal elements of the density matrix. Conversely the von Neumann's entropy is sensitive to all information in the system, stationary or not. This sensitivity results in $S_n$ being time independent for any dynamical processes in a closed system, because time evolution is a unitary transformation, which does not change the total information content.

\acknowledgements The author acknowledges helpful discussions with R. Barankov on earlier stages of this work. The author also thanks C. Gogolin for sharing the proof of Eq.~(\ref{second_law_sum}) and for many valuable comments. It is also a pleasure to acknowledge E.~Altman, S.~Girvin, V.~Gritsev, V.~Gurarie, D.~Huse, Y.~Kafri, W.~Zwerger for helpful discussions related to this work. This work was supported by NSF (DMR-0907039), AFOSR YIP, AFOSR FA9550-10-1-0110, and Sloan Foundation.

\end{document}